\documentstyle[preprint,aps,prl]{revtex}
\begin{document}
\draft
\title
{Variable-phase method and Levinson's theorem in two dimensions:
Application to a screened Coulomb potential}

\author{M. E. Portnoi\cite{byline} and I. Galbraith} 
\address{
Physics Department, Heriot-Watt University, Edinburgh EH14 4AS, 
United Kingdom} 

\date{12 February 1997}

\maketitle

\begin{abstract}
The variable-phase approach is applied to scattering and bound states in 
an attractive Coulomb potential, statically screened by a two-dimensional (2D) 
electron gas. A 2D formulation of Levinson's theorem is used for 
bound-state counting and  a hitherto undiscovered, simple relationship
between the screening length and the number of bound states  is found.
As the screening length  is increased, sets of 
bound states with differing quantum numbers appear degenerately.
\end{abstract}



The quantum mechanics of low-dimensional systems has become a major 
research field with the advent of growth techniques for the realization 
of semiconductor quantum wells. Almost all of the computational techniques 
developed for three-dimensional (3D) problems have already been 
extended to lower dimensions. 
The variable-phase method for the calculation of the scattering phase 
shifts was introduced long ago by Morse and Allis \cite{MA33} and 
expanded by Calogero \cite{Cal67}, Babikov \cite{Babikov} and 
others \cite{VPA} but has to our knowledge never been used for analysis 
of any realistic two-dimensional problem.
In this Letter we  use the variable-phase method to treat both scattering 
and bound states on the same footing for the Coulomb potential statically 
screened by 2D electron gas. This potential plays a central role in the 
physics of semiconductor heterostructures, one of the most rapidly growing 
fields in condensed matter physics. Despite having been studied for 
approximately 30 years \cite{Stern,SH67} this potential exhibits
some peculiar features which have never been noticed until now, e.g., 
as the screening length increases, sets of bound states with differing 
quantum numbers appear degenerately. These degeneracies are not, 
however, the same as the degeneracies in the unscreened spectrum. 
We also report a hitherto undiscovered, simple relationship connecting 
the number of bound states to the screening length. 

In what follows we consider a simple case in which a screened charge 
resides in the same plane as a screening 2D electron gas. This geometry 
is appropriate to the problem of screened excitons \cite{HKbook} 
or impurities \cite {SH67} in a narrow quantum well. 
We assume that an attractive potential is created by a point charge 
$e$ at the origin and use throughout this paper excitonic Rydberg units 
where length and energy are scaled, respectively, by the effective Bohr 
radius $~a^*~$ and Rydberg $~Ry^*~$. In these units and geometry 
the Thomas-Fermi expression for the statically screened potential 
in the electron plane \cite{SH67,HKbook} is 
\begin {equation} 
V_s(q) ~=~ - 2 {2 \pi \over {q+q_s}}~, 
\label{eq1} 
\end {equation} 
where $~q_s~$ is the 2D screening wavenumber. Eq.~(\ref{eq1}) is the 2D   
analogue of the Yukawa potential. Taking the 2D Fourier transformation 
of Eq.~(\ref{eq1}) yields in real space 
\begin {equation} 
V_s(\rho) ~=~ -2 \left \{{1 \over \rho} - q_s \int_{0}^{\infty} 
{J_{0}(q \rho) \over {q+q_s}} dq \right \}~, 
\label{eq2}
\end {equation} 
where $~J_{0}(q\rho)~$ is the Bessel function, and  
$~\rho = (x^2 + y^2)^{1/2}~$ is the in-plane distance from the origin. 

To describe the application of the variable-phase method in 2D we 
consider a particle moving with energy $~E=k^2~$ in the potential  
$~V(\rho)~$ which has  radial symmetry.  Since the potential is symmetric,  
we can separate variables in the expression for the wave function: 
\begin {equation}
\Psi(\rho,\varphi)~=~\sum_{m=0}^{\infty}R_m(\rho)~\cos(m\varphi)~,~  
\label{eq8}
\end {equation}
where $m$ is the absolute value of the projection of the angular momentum 
onto the symmetry axis of the potential. 
At large distances from the scattering center, the radial function satisfies 
the free Bessel equation, whose general solution is 
\begin {eqnarray} 
R_m(\rho)&&~=~A_m[J_m(k\rho) \cos{\delta_m}-N_m(k\rho) \sin{\delta_m}]~ 
\nonumber \\ 
&&\stackrel{\rho\rightarrow\infty}{\longrightarrow}~ 
A_m\left({2\over\pi k \rho}\right)^{1/2} 
\cos(k\rho - (2m+1)\pi/4 + \delta_m)~,~ 
\label{eq9} 
\end{eqnarray} 
where $~\delta_m~$ is the scattering phase shift \cite{SH67,Ian84},  
$~J_m(k\rho)~$ and $~N_m(k\rho)~$ are the Bessel and Neumann functions,  
respectively. Both total and transport cross sections in 2D can be 
expressed  via the scattering phase shifts in a simple fashion 
\cite{SH67}. 

In the variable-phase method $~A_m~$ and $~\delta_m~$ are considered 
not as constants but as functions of the distance $~\rho~$. 
The phase function $\delta_m(\rho)$ is the phase shift produced by  
a potential cut-off at a distance $\rho$. 
Then the scattering phase shift can be obtained as a large distance 
limit  of the phase function $~\delta_m(\rho)~$, which satisfies 
the following first-order, non-linear differential equation 
 originating from the radial Schr\"{o}dinger equation\cite{Babikov}: 
\begin {equation} 
{d \over d\rho} \delta_m(\rho)~=~-{\pi \over 2}\rho V(\rho) 
[J_m(k\rho)\cos{\delta_m(\rho)} - N_m(k\rho)\sin{\delta_m(\rho)}]^2 
\label{eq10} 
\end {equation} 
with the boundary condition 
\begin {equation} 
\delta_m(0)~=~0~. 
\label{eq11} 
\end {equation} 
Eq.~(\ref{eq11}) ensures that the radial function does not diverge at 
$~\rho=0~$. 
Physically one can view the phase function as measuring the retardation 
of the scattering  wave function due to the potential. From this view it 
is clear that Eq.~(\ref{eq11}) is the correct boundary condition since 
the potential can produce no retardation at the origin. The total scattering 
phase shift is given as an asymptote of $~\delta_m(\rho)~$: 

\begin {equation}
\delta_m~=~\lim_{\rho \rightarrow \infty}\delta_m(\rho)~.
\label{eq12}
\end {equation}
For numerical convenience, instead of the boundary condition 
Eq.~(\ref{eq11}), the small-$\rho~$ expansion is used 
\begin {equation} 
\delta_m(\rho)~\approx~-{\pi k^{2m} \over 2^{2m+1} (m!)^2} 
{\int}_{0}^{\rho}V(\rho^\prime){\rho^\prime}^{2m+1}d\rho^\prime~, 
~~~\rho~\rightarrow~0~. 
\label{eq13}
\end {equation}
It is  also useful to rewrite Eq.~(\ref{eq2}) in terms of special 
functions \cite{SH67,GRyz} as 
\begin {equation} 
V_s(\rho) ~=~ -2 \left \{{1 \over \rho}-  
{\pi \over 2} q_s [{\bf H}_{0}(q_s \rho)  
- N_{0}(q_s \rho)] \right\}~, 
\label{eq3}
\end {equation} 
where $~{\bf H}_{0}(q_s \rho)~$ and $~N_{0}(q_s \rho)~$ are the Struve and 
Neumann functions, respectively. Series representations of these functions 
for small values of argument and asymptotic expansions for large values of 
argument allow the accurate calculation of $~V_s(\rho)~$ for all 
values of $~q_s \rho~$. Asymptotic expressions for the potential are 
\begin {equation} 
V_s(\rho)~\sim~-{2 \over q_s^2 \rho^3} \left 
\{1+\sum_{n=1}^{p}~(-1)^n~{[(2n+1)!!]^2 \over (q_s \rho)^{2n}} 
~+~O[(q_s \rho)^{-2p-2}]~\right \}~,~~~ q_s \rho ~\gg~1~,~ 
\label{eq4} 
\end {equation} 
and 
\begin {equation}
V_s(\rho)~\sim~-2[\rho^{-1}~+~q_s\ln(Cq_s \rho)]+O(q_s \rho)~, 
~~~q_s \rho~\ll~1~, 
\label{eq5} 
\end {equation} 
where $~C={1 \over 2} \exp(\gamma)~$, and $~\gamma~$ is Euler's constant. 
The asymptotic expressions Eqs.~(\ref{eq4},\ref{eq5}) show  that the 
potential $~V_s(\rho)~$ satisfies the conditions: 
\begin {equation} 
{\int}_{\rho}^{\infty}V(\rho^\prime)~d\rho^\prime~\rightarrow~0~, 
~~~\rho~\rightarrow~\infty~,~ 
\label{eq7} 
\end {equation} 
and 
\begin {equation} 
{\rho}^{2} V(\rho)~\rightarrow~0~, ~~~\rho~\rightarrow~0~,~ 
\label{eq6} 
\end {equation} 
which are sufficient to allow application of the variable-phase method 
for this potential. 

\begin{figure}
\begin{center}
\includegraphics{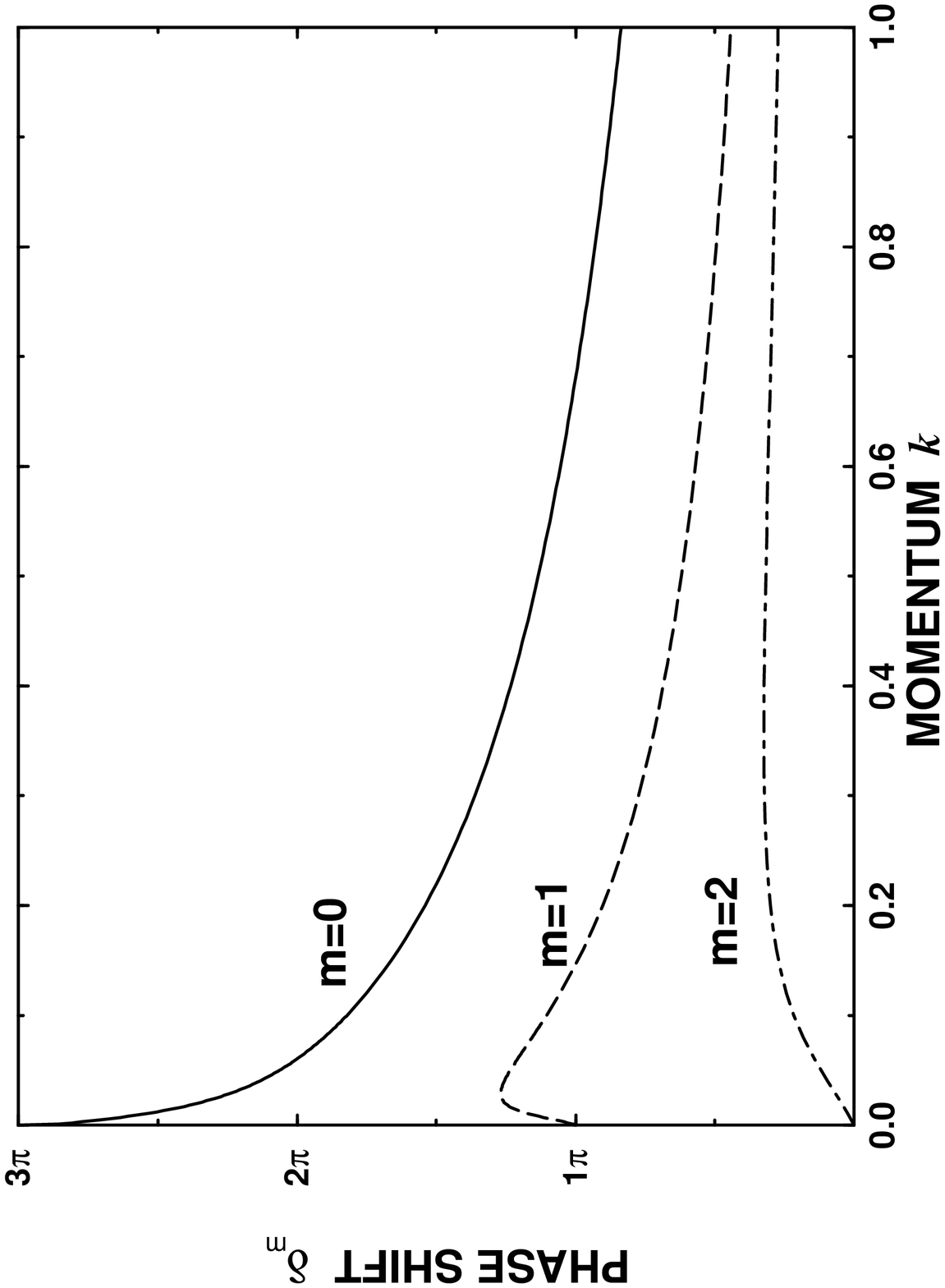}
\vskip 16truecm
\end{center}
\caption{Scattering phase shifts versus in-plane wave vector $~k~$ 
(in units of inverse Bohr radius $~1/a^*~$) for a 2D particle in 
an attractive Coulomb potential screened by a 2D electron gas,  
screening length $~r_s=1/q_s=5a^*~$. Numbers show $~m~$ values.}
\label{fig1}
\end{figure}

Figure \ref{fig1} shows the $~k~$ dependence of the phase shifts $~\delta_m~$ 
obtained by the numerical solution of Eq.~(\ref{eq10}) with the initial 
condition  Eq.~(\ref{eq13}) for the screened Coulomb potential $~V_s(\rho)~$.  
The distinctive feature of this plot to which we draw attention is that 
in the low-energy limit, $~k \rightarrow 0~$, the scattering phase shift 
is an integer number of $~\pi~$: 
\begin {equation}
\lim_{k \rightarrow 0} \delta_m~=~\nu\pi~. 
\label{eq14}
\end {equation}
This behavior may be understood recalling Levinson's theorem 
\cite{Levinson} which connects the zero-energy scattering phase shift 
with the number of the bound states for non-relativistic particles in 3D. 
Recently Levinson's theorem has been discussed  for Dirac particles, 
multichannel scattering, multi-particle single-channel scattering, 
one-dimensional scattering systems, systems with non-uniform effective mass, 
and even for time-periodic potentials \cite{NewLevinson}. 
However its applicability to the 2D scattering problem has not been 
considered yet. We expect that this fundamental theorem holds also in 2D 
in the form of Eq.~(\ref{eq14}), where $~\nu~$ is the number of bound states 
for a given $m$. A simple proof for $~m \ne 0~$ follows from Eq.~(\ref{eq10}) 
in the same fashion as Calogero's proof of Levinson's theorem in the 
3D case \cite{Cal67}. For $~m=0~$ the proof is more complicated due to 
the logarithmic divergence of the Neumann function,$~N_0(k\rho)~$. 
A rigorous proof of  Levinson's theorem in 2D, based on analytic 
properties of the scattering matrix \cite{Sitenko}, is needed. 
We have, however, verified the validity of Eq.~(\ref{eq14}) for 
{\em all} $~m~$ values in the analytically tractable case of a circular 
finite potential well\cite{Portnoi88}. 
     
Despite its appeal, Levinson's theorem has not been widely used to enumerate 
bound states since there exists an ambiguity in the usual definition 
of $\delta_m$, being defined only up to $~{\rm mod}(\pi)~$\cite{SchSchw65}. 
However using the variable phase approach avoids this problem since the phase 
function is uniquely defined by Eqs. (\ref{eq10},\ref{eq11}) for all $\rho$. 
  
\begin{figure}
\begin{center}
\includegraphics{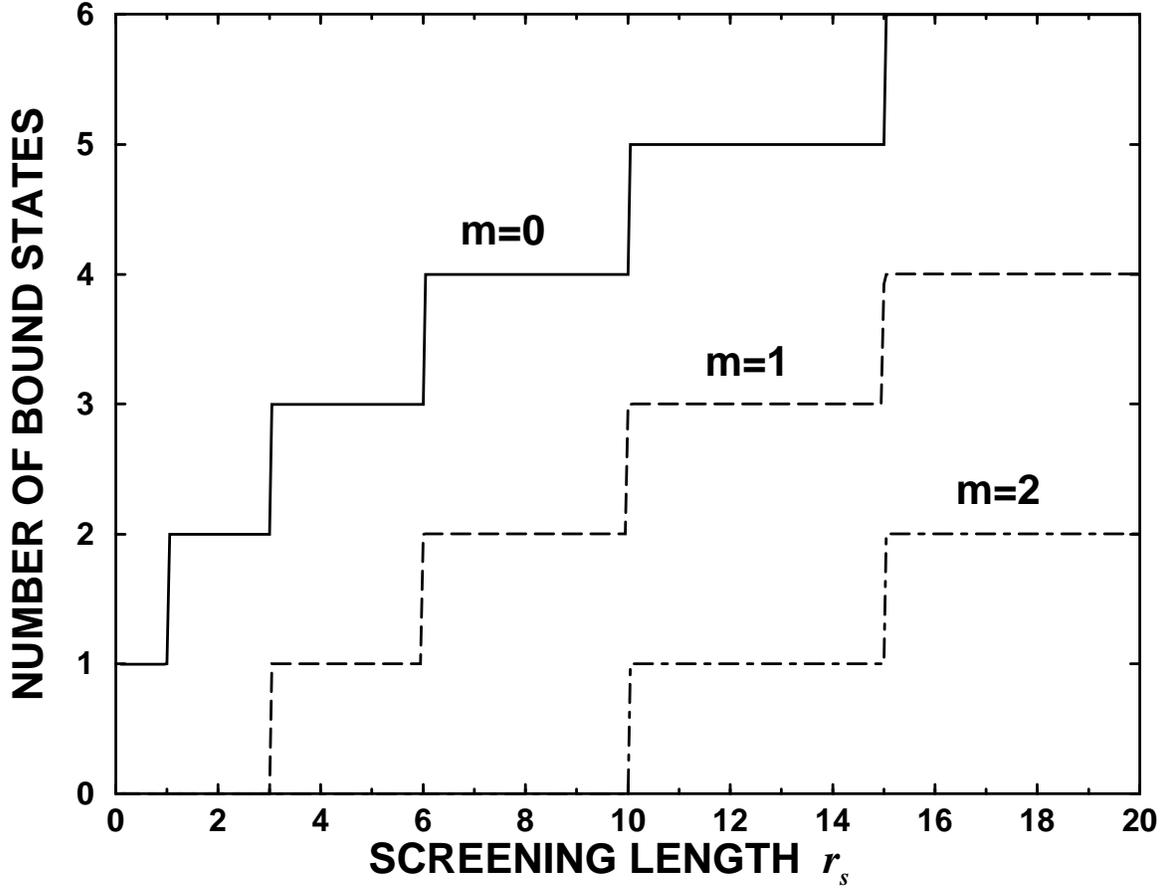}
\end{center}
\vskip 16truecm
\caption{
Number of bound states, calculated as the low-energy limit of the phase 
shift, plotted versus the screening length $~{r_s}~$ 
(in the units of Bohr radius $~a^*~$).  The solid line corresponds 
to $~m=0~$; dashed line, $~m=1~$; dashed-dotted line, $~m=2~$.  
New steps appear at $~{(r_s)}_c~=~(2m+\nu-1)(2m+\nu)/2~$, 
$~\nu~=~1,~2,~\ldots~$.
}
\label{fig2}
\end{figure}

In Fig.~\ref{fig2} the number of bound states, obtained as the low-energy 
limit of the scattering phase shift in units of $~\pi~$, is plotted as 
a function of the screening length $~r_s=1/q_s~$ for the potential $~V_s~$. 
As the screening length increases, the potential supports more bound states 
and these new bound states appear at critical values of the screening length 
indicated by the steps. One  can see from the location of these steps that 
the $\nu$-th bound state for a given $m$  appears at the  critical screening 
length, given by a simple formula 
\begin {equation} 
{(r_s)}_c~=~{(2m+\nu-1)(2m+\nu) \over 2}~,~~~~\nu=1,2,~\ldots~. 
\label{eq15}  
\end {equation} 
This intriguingly simple relation has, to our knowledge, never been reported  
despite numerous calculations of the binding energy in the screened Coulomb 
potential, since conventional numerical methods for the binding energy 
calculation fail for extremely shallow energy levels. Note that for $m=0$ 
the first bound state appears immediately at $ {(r_s)}_c~=~0$, corresponding 
to the fact that there is always at least one bound state in any symmetric 2D 
attractive potential. Using Eq.~(\ref{eq15}) we can simply evaluate how many 
bound states the 2D statically-screened Coulomb potential will support 
for any value of $m$. 
 
The  Bargmann bound condition \cite{Bargmann} re-stated for two dimensions 
\cite{SH67} is  $~m\nu<r_s~$. This gives  a gross over-estimation of the 
number of the bound states; e.g., for $~r_s=4,~m=1~$ this implies only that 
there are less than four bound states whereas there is in fact only one. 
In this sense the Bargmann condition is of limited utility. 

For many applications (e.g., in a partition function calculation) the value 
of the binding energy, not just the number of bound states, is important. 
The variable-phase method provides an elegant and efficient solution of the 
eigenvalue problem as well. To approach this problem we recall that for the 
states with  negative energy the wavenumber $~k~$ is imaginary, 
$~k=i \kappa~$, and we introduce the function $~\mu_m (\rho,\kappa)~$ 
vanishing in the origin and satisfying a non-linear equation \cite{Babikov} 
\begin {equation} 
{d \over d\rho} \mu_m(\rho,\kappa)~=~-{\pi \over 2}\rho V(\rho) 
\left[ I_m(\kappa \rho)\cos{\mu_m(\rho,\kappa)}+ 
{2 \over \pi} K_m(\kappa \rho)\sin{\mu_m(\rho,\kappa)}\right]^2~,~~ 
\label{eq16}
\end {equation}
where $~I_m(\kappa \rho)~$ and $~K_m(\kappa \rho)~$ are the modified Bessel 
functions of the first and  second kind, respectively. The functions 
$~I_m(\kappa \rho)~$ and $~K_m(\kappa \rho)~$ represent two linearly 
independent solutions of the free radial-wave Schr\"{o}dinger equation 
for the negative value of energy, $~E=-\kappa^2~$, and $~\cot{\mu_m}~$ 
characterizes the weights of the diverging ($I_m$) and converging ($K_m$) 
solutions as $~\rho \rightarrow \infty~$. For the bound state, the 
diverging solution vanishes, implying the asymptotic condition 
\begin {equation}
\mu_m(\rho \rightarrow \infty~,\kappa_\nu)~=~(\nu-1/2)\pi~,
~~~~\nu=1,2,~\ldots~.
\label{eq17} 
\end {equation} 
Here $\nu$ numerates the bound states for a given $m$ and $~(\nu-1)~$ is the 
number of non-zero nodes of the radial wave function. For numerical solution 
of Eq.~(\ref{eq16}) instead of the boundary condition $~\mu_m(0,\kappa)=0~$ 
an approximate initial condition (analogous to the condition Eq.~(\ref{eq13}) 
for the phase function $~\delta_m(\rho)~$) is used. 

\begin{figure}
\begin{center}
\includegraphics{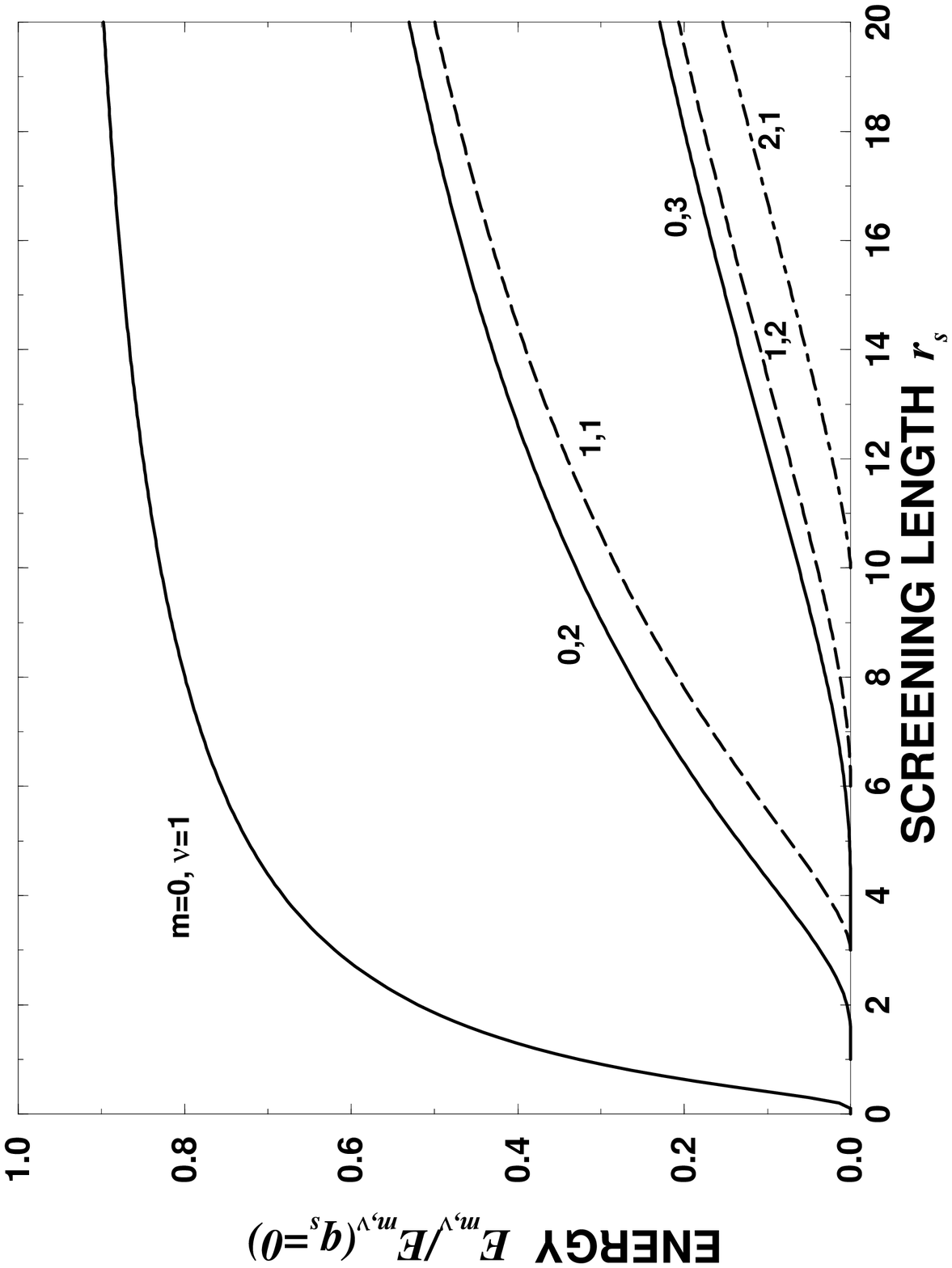}
\end{center}
\vskip 16truecm
\caption{
The bound-state energies $~E_{m,\nu}~$ plotted versus the screening length 
$~{r_s}~$ for different $~m~$ values. The curves are normalized 
by the energies for the unscreened potential, $~E_{m,\nu}(q_s=0)=-(m+\nu-1/2)^{-2}~$. 
Numbers show $~m~$ and $~\nu~$ values. 
}
\label{fig3}
\end{figure}

The bound-state energies $~E_{m,\nu}~$ versus the screening length $~r_s~$ for  
the potential $~V_s~$ are plotted in Fig.~\ref{fig3}. For clarity all the 
curves are normalized by the values of the energy for the unscreened potential 
\cite{Flugge}, $~E_{m,\nu}(q_s=0)=-(\nu+m-1/2)^{-2}=-(n+1/2)^{-2}~$. 
The principal quantum number $~n~$ is given by $~n=\nu+m-1~$. We note that 
the unscreened eigenstates with the same value of $~m+\nu~$  are degenerate. 
A consequence of the relationship in Eq.~(\ref{eq15}) is that when the bound 
states first appear (at $~r_s={(r_s)}_c~$), eigenstates with the same value 
of $~2m+\nu~$ are degenerate. With increasing $~r_s~$ this degeneracy is  
lifted and the degeneracy for the states with the same values of $~n~$ 
appears as $~r_s \rightarrow \infty~$. 

In conclusion, we have used the variable-phase method to study scattering  
and bound states in two dimensions. The 2D analogy of Levinson's theorem 
is formulated and verified empirically. 
A hitherto undiscovered, simple relationship between the screening length 
and the number of bound states in the statically-screened Coulomb potential  
has been found. Challenging problems such as a general proof of Levinson's  
theorem in 2D and the analytic derivation of Eq.~(\ref{eq15}) are clearly  
the subject of further research. 

This work was supported by the Engineering and Physical Science Research 
Council and the Royal Society.

\newpage

\newpage

\begin{references}

\bibitem[*]{byline} Also at A. F. Ioffe Physico-Technical Institute, 194021 
St.Petersburg, Russia.
\bibitem{MA33} P. M. Morse and W. P. Allis, Phys. Rev. {\bf 44}, 269 (1933).
\bibitem{Cal67} F. Calogero, {\it Variable Phase Approach to Potential Scattering} (Academic, New York, 1967);
F. Calogero, Nuovo Cimento {\bf 27}, 261 (1963).
\bibitem{Babikov} V. V. Babikov, Usp. Fiz. Nauk {\bf 92}, 3 (1967)  
[Sov. Phys. Usp. {\bf 92}, 271 (1967)];
V. V. Babikov, {\it Method of Phase Functions in Quantum Mechanics} (Nauka, 
Moscow, 1976) [in Russian].  
\bibitem{VPA} B. R. Levy and J. B. Keller, J. Math. Phys. {\bf 4}, 54 (1963); 
A. Degasperis, Nuovo Cimento {\bf 34}, 1667 (1964); 
J. R. Cox, {\it ibid.} {\bf 37}, 474 (1965);
J. M. Clifton and R. A. Leacock, J. Comput. Phys. {\bf 38}, 327 (1980);
U. Das, J. Math. Phys. {\bf 22}, 1045 (1981) and references therein.   
\bibitem{Stern} F. Stern, Phys. Rev. Lett. {\bf 18}, 546 (1967);
for the review of later works see T. Ando, A. B. Fowler, and F. Stern, 
Rev. Mod. Phys. {\bf 54}, 437 (1982).
\bibitem{SH67} F. Stern and W. E.  Howard,  Phys. Rev. {\bf 163}, 816 (1967).
\bibitem{HKbook} H. Haug and S. W. Koch, {\it Quantum Theory of the Optical 
and Electronic Properties of Semiconductors} 
(World Scientific, Singapore, 1994).  
\bibitem{Ian84} I. Galbraith, Y. S. Ching, and E. Abraham, Am. J. Phys. 
{\bf 52}, 60 (1984).
\bibitem{GRyz} I. S. Gradshteyn and I. M. Ryzhik, {\it Table of Integrals,
 Series and Products} (Academic, New York, 1980).
\bibitem{Levinson} N. Levinson, K. Dans. Vidensk. Selsk., Mat. Fys. Medd. 
{\bf 25}, 3 (1949).  
\bibitem{NewLevinson} 
Zh.-Q. Ma, Phys. Rev. Lett. {\bf 76}, 3654 (1996); 
N. Poliatzky, {\it ibid.} {\bf 76}, 3655 (1996);  
F. Vidal and J. Letourneaux, Phys. Rev. C {\bf 45}, 418 (1992); 
L. Rosenberg and L. Spruch, Phys. Rev. A {\bf 54}, 4978 (1996) 
and {\it ibid.} {\bf 54}, 4985 (1996); 
K. A. Kiers, W. van Dijk, J. Math. Phys. {\bf 37}, 6033 (1996);  
M. S. Debianchi, {\it ibid.} {\bf 35}, 2719 (1994); 
V. Milanovi\'{c} {\it et al.}, Phys. Lett. A {\bf 170}, 127 (1992); 
V. Milanovi\'{c} {\it et al.}, J. Phys. A {\bf 25}, L1305 (1992); 
P. A. Martin and M. S. Debianchi, Europhys. Lett. {\bf 34}, 639 (1996)   
and references therein.
\bibitem{Portnoi88} M. E. Portnoi, Pis'ma Zh. Tekh. Fiz. {\bf 14}, 1252 (1988) 
[Sov. Tech. Phys. Lett. {\bf 14}, 547 (1988)].   
\bibitem{Sitenko} A. G. Sitenko, {\it Lectures in Scattering Theory} 
(Pergamon, Oxford, 1971).  
\bibitem{SchSchw65} H. M. Schey and J. L. Schwartz, Phys. Rev. {\bf 139}, 
1428 (1965).
\bibitem{Bargmann} V. Bargmann, Proc. Nat. Acad. Sc. USA, {\bf 38}, 961 (1952);
see also M. Reed and B. Simon, {\it Analysis of Operators} 
(Academic, New York, 1978).
\bibitem{Flugge} S. Fl\"{u}gge and H. Marschall, {\it Rechenmethoden der 
Quantentheorie} (Springer, Berlin, 1952), Problem 24.

\end{references}
\end{document}